\documentclass[prd,showpacs,letterpaper,twocolumn,nofootinbib,longbibliography,preprintnumbers,floatfix,superscriptaddress]{revtex4-2}

\usepackage{CJKutf8}
\usepackage{tikz}
\usepackage{adjustbox}
\usepackage{amsmath}
\usepackage{graphicx}
\usepackage{fullpage}
\usepackage[colorlinks=true,allcolors=blue]{hyperref}
\usepackage[capitalise]{cleveref}
\usepackage[utf8]{inputenc}
\usepackage{siunitx}
\usepackage{array}
\DeclareSIUnit \s {\second}
\DeclareSIUnit \ns {\nano\second}
\DeclareSIUnit \mus {\micro\second}
\DeclareSIUnit \ms {\milli\second}
\DeclareSIUnit \MB {\mega\byte}
\DeclareSIUnit \GB {\giga\byte}
\DeclareSIUnit \TB {\tera\byte}
\DeclareSIUnit \PB {\peta\byte}
\DeclareSIUnit \Mbps {\mega\bit/\s}
\DeclareSIUnit \Gbps {\giga\bit/\s}
\DeclareSIUnit \Tbps {\tera\bit/\s}
\DeclareSIUnit \Pbps {\peta\bit/\s}
\DeclareSIUnit \kton {\kilo\tonne} % changed  back to kton
\DeclareSIUnit \kt {\kilo\tonne}
\DeclareSIUnit \Mt {\mega\tonne}
\DeclareSIUnit \eV {\electronvolt}
\DeclareSIUnit \keV {\kilo\electronvolt}
\DeclareSIUnit \MeV {\mega\electronvolt}
\DeclareSIUnit \GeV {\giga\electronvolt}
\DeclareSIUnit \TeV {\tera\electronvolt}
\DeclareSIUnit \PeV {\peta\electronvolt}
\DeclareSIUnit \EeV {\exa\electronvolt}
\DeclareSIUnit \m {\meter}
\DeclareSIUnit \cm {\centi\meter}
\DeclareSIUnit \in {\inchcommand}
\DeclareSIUnit \km {\kilo\meter}
\DeclareSIUnit \kV {\kilo\volt}
\DeclareSIUnit \kW {\kilo\watt}
\DeclareSIUnit \MW {\mega\watt}
\DeclareSIUnit \MHz {\mega\hertz}
\DeclareSIUnit \mrad {\milli\radian}
\DeclareSIUnit \year {years}
\DeclareSIUnit \POT {POT}
\DeclareSIUnit \sig {$\sigma$}
\DeclareSIUnit\parsec{pc}
\DeclareSIUnit\lightyear{ly}
\DeclareSIUnit\foot{ft}
\DeclareSIUnit\ft{ft}
\DeclareSIUnit \ppb{ppb}
\DeclareSIUnit \ppt{ppt}
\DeclareSIUnit \samples{S}
\DeclareSIUnit \pe{PE}

\newcommand{\enu}{\E_\enu}

\usepackage{pifont}
\usepackage{enumitem}
\usepackage{url}
\usepackage{rotating}
\usepackage{epsfig,graphics,rotate}
\usepackage{flushend}
\usepackage{bm}
\usepackage{amssymb}
\usepackage{amsmath}
\usepackage{amsfonts}
\usepackage{gensymb}
\usepackage{booktabs}

\usepackage{microtype}
\usepackage{multirow, makecell}
\usepackage{lipsum}

\usepackage{xspace}
\usepackage{comment}
\usepackage{floatrow}
\usepackage{ragged2e}

\usepackage{nicefrac, xfrac}
\usepackage{mathtools} % for 'pmatrix' environment
\usepackage{relsize}   % for '\larger' macro

\usepackage{float}

\definecolor{lime}{HTML}{A6CE39}
\DeclareRobustCommand{\orcidicon}{
	\begin{tikzpicture}
	\draw[lime, fill=lime] (0,0) 
	circle [radius=0.16] 
	node[white] {{\fontfamily{qag}\selectfont \tiny ID}};
	\draw[white, fill=white] (-0.0665,0.095) 
	circle [radius=0.005];
	\end{tikzpicture}
	\hspace{-2mm}
}

\foreach \x in {A, ..., Z}{\expandafter\xdef\csname orcid\x\endcsname{\noexpand\href{https://orcid.org/\csname orcidauthor\x\endcsname}
			{\noexpand\orcidicon}}
   }

\begin{document}

%\preprint{IPPP/24/05}

\begin{CJK*}{UTF8}{gbsn}
\title{Boosting Neutrino Mass Ordering Sensitivity with Inelasticity for Atmospheric Neutrino Oscillation Measurement}
\author{Santiago Giner Olavarrieta\orcidE{}}
\email{santiagoginer@college.harvard.edu}
\affiliation{Department of Physics \& Laboratory for Particle Physics and Cosmology, Harvard University, Cambridge, MA 02138, USA}

\author{Miaochen Jin (靳淼辰)\orcidD{}}
\email{miaochenjin@g.harvard.edu}
\affiliation{Department of Physics \& Laboratory for Particle Physics and Cosmology, Harvard University, Cambridge, MA 02138, USA}

\author{\\Carlos A.~Arg{\"u}elles\orcidA{}}
\email{carguelles@fas.harvard.edu}
\affiliation{Department of Physics \& Laboratory for Particle Physics and Cosmology, Harvard University, Cambridge, MA 02138, USA}

\author{Pablo Fern\'andez\orcidB{}}
\email{pablo.fernandez@dipc.org}
\affiliation{Donostia International Physics Center DIPC, San Sebasti\'an/Donostia, E-20018, Spain}

\author{Ivan Martinez-Soler\orcidC{}}
\email{ivan.j.martinez-soler@durham.ac.uk}
\affiliation{Department of Physics \& Institute for Particle Physics Phenomenology, University of Durham, Durham, DH1 3LE, United Kingdom}
\date{\today}

\begin{abstract}
In this letter, we study the potential of boosting the atmospheric neutrino experiments sensitivity to the neutrino mass ordering (NMO) sensitivity by incorporating inelasticity measurements.
We show how this observable improves the sensitivity to the NMO and the precision of other neutrino oscillation parameters relevant to atmospheric neutrinos, specifically in the IceCube-Upgrade and KM3NeT-ORCA detectors.
Our results indicate that an oscillation analysis of atmospheric neutrinos including inelasticity information has the potential to enhance the ordering discrimination by several units of $\chi^2$ in the assumed scenario of 5 and 3 years of running of IceCube-Upgrade and KM3NeT-ORCA detectors, respectively.
\end{abstract}
\maketitle
\end{CJK*}

\textbf{\emph{Introduction---}}
%\label{sec:intro}
Unmagnetized neutrino experiments cannot distinguish neutrinos from anti-neutrinos on an event-by-event basis when studying neutrino-nucleon interactions.
Instead, experiments operating without a magnet need to rely on the particle content of the beam or the use of kinematic observables to statistically separate neutrinos from antineutrinos.

So far, analysis of atmospheric neutrinos aimed to determine neutrino oscillation parameters~\cite{Super-Kamiokande:2017yvm,ANTARES:2018rtf,Pestel:2022ewd,IceCubeCollaboration:2023wtb} has not exploited kinematic variables aimed at distinguishing between neutrinos and antineutrinos in large telescopes based on water/ice-Cherenkov.

This missing information undoubtedly hides the potential of atmospheric neutrinos in measuring the different oscillation effects that are distinct between neutrinos and antineutrinos, namely the mass ordering through the Mikheyev-Smirnov-Wolfenstein (MSW)~\cite{Wolfenstein:1977ue,Mikheyev:1985zog} Earth matter effects and the magnitude of the $CP$-violating phase in the lepton sector~\cite{Razzaque:2014vba}.
This is particularly important since, as recently demonstrated in Ref.~\cite{Arguelles:2022hrt}, atmospheric neutrino experiments are expected to yield the most precise measurements of the neutrino oscillation parameters and are expected to determine the neutrino mass ordering by the end of the decade.

With this motivation, the Super-Kamiokande experiment has already implemented various techniques to distinguish neutrinos from antineutrinos~\cite{Super-Kamiokande:2022hxq}: detecting low-energy secondary particles like Michel electrons or neutrons, and computing kinematically relevant variables when possible, that is in the multi-ring samples.
Producing a first neutrino oscillation analysis with neutrino-anti-neutrino event based on neutron tagging~\cite{Super-Kamiokande:2023ahc}.

\begin{figure}[b]
    \centering
    \includegraphics[width=\textwidth]{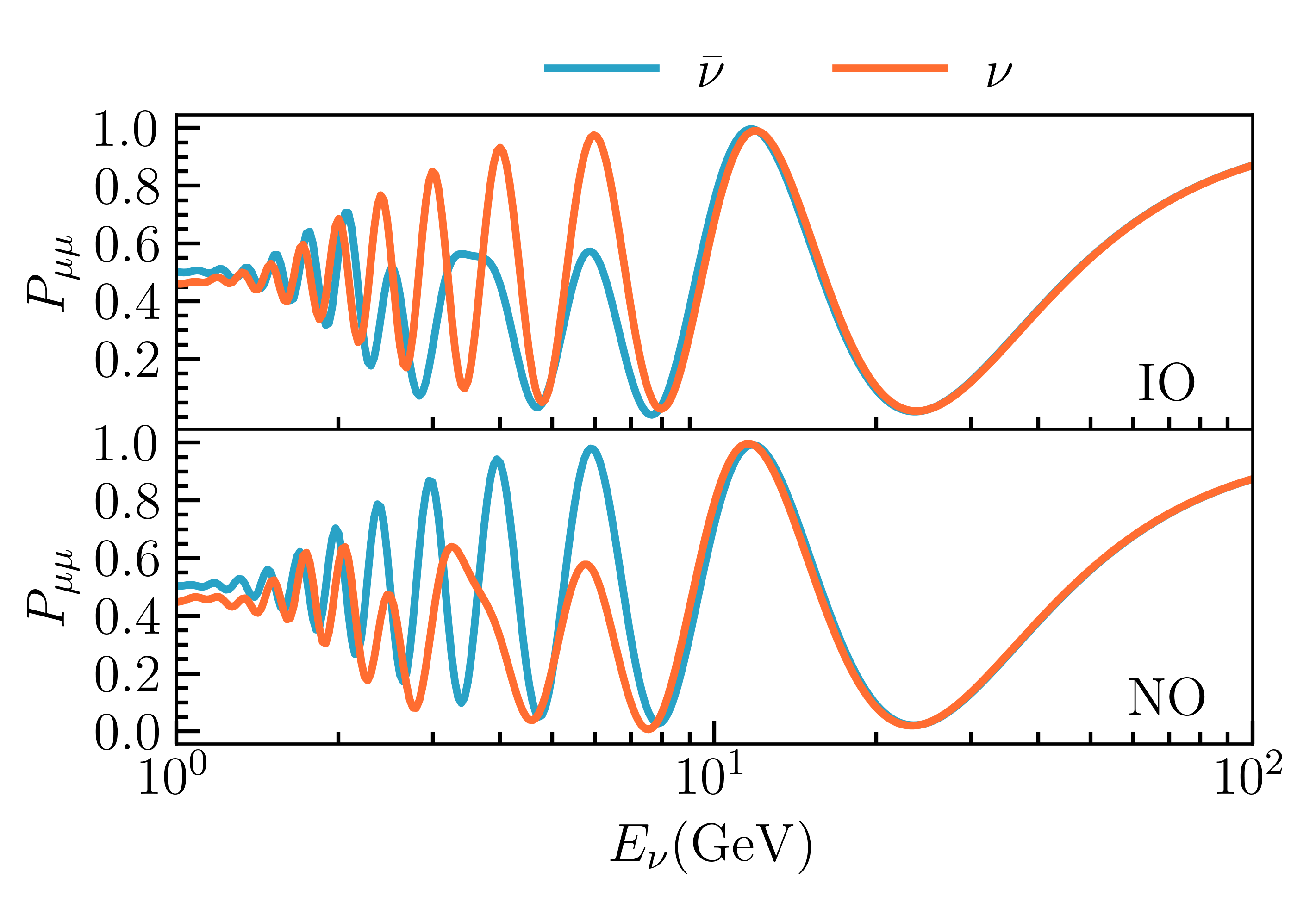}
    \caption{Muon disappearance probabilities for both mass orderings, normal (NO) and inverted (IO), and both neutrinos and anti-neutrino, considering a trajectory crossing the entire Earth ($\cos\theta_{\nu} = -0.95$). As a best fit parameters, we used the results of latest global analysis~\cite{Esteban:2020cvm}. A Gaussian filter with a width of $5\%\sqrt{E_{\nu}}$ has been included to remove the fast oscillations at lower energies.} 
    \label{fig:oscillations}
\end{figure}

The former requires a low-background detector with very high photo-sensor coverage, which is currently out of reach for the large, next-generation, multi-megatonne neutrino detectors.
On the other hand, in this work, we demonstrate how the IceCube-Upgrade and KM3NeT-ORCA detectors could use the reconstructed inelasticity of the neutrino interaction to improve the sensitivity to neutrino mass ordering and the $CP$ phase.
The inelasticity $y$, also known as Bjorken-$y$, is the fraction of the neutrino energy transferred to a hadronic system with which the neutrino interacts. There are some efforts that has already been taken in the IceCube collaboration in reconstructing the inelasticity of low energy events~\cite{Peterson:2023ayg}.
Thus, in this letter, we extend the work in Ref.~\cite{Arguelles:2022hrt} by studying the impact in sensitivity of the IceCube-Upgrade and KM3NeT-ORCA neutrino telescopes that comes from incorporating an event's inelasticity in the oscillation analysis. 

The results obtained in this work further complement the motivation of a combined oscillation analysis of atmospheric neutrinos to provide a precise picture of the mixing scenario independent from the current and early measurements of the next-generation accelerator experiments.
As well as motivate the development of techniques that enable the reconstruction of the inelasticity in neutrino telescopes.

%This letter is structured as follows: In \Cref{sec:atm_nu}, we motivate and describe the expected difference in oscillations between neutrinos and antineutrinos and the effect of their separation in the sensitivity to the oscillation parameters.
%In~\cref{sec:experiments}, we briefly describe the experiments considered in this work, IceCube-Upgrade and ORCA.
%The simulation of both experiment data sets and the methods developed to generate the reconstructed inelasticity based on each detector's performance is described in~\cref{sec:methods}.
%In~\cref{sec:results}, we explain how the inelasticity is included in the combined statistical analysis and show the main results from this analysis.
%We finally conclude and discuss the results in~\cref{sec:conclusion}.

\textbf{\emph{Atmospheric neutrinos and antineutrinos ---}} 
%\label{sec:atm_nu}
The study of neutrino oscillations has entered an era of high precision, where only a few aspects remain unknown.
Among these unknowns are the octant of the $\theta_{23}$, the mass ordering, and the CP-phase, $\delta_{CP}$.
Both the mass ordering and $\delta_{CP}$ predict different behaviors for neutrinos and anti-neutrinos as they propagate through Earth.
Specifically, in the case of normal mass ordering (NO), where $m_{3}> m_{2}, m_{1}$, a matter-induced resonance is predicted for neutrinos crossing the mantle and the core of the Earth at energies around $6$~GeV.
In the case of inverted ordering (IO), where $m_{3} < m_{1}, m_{2}$, this resonance occurs in the anti-neutrino propagation, as illustrated in~\Cref{fig:oscillations}.
A similar situation arises in the case of $\delta_{CP}$, where, in the presence of $CP$-violation, the oscillation evolution differs between neutrinos and antineutrinos.
For a detailed description of the effects of $CP$-violation on atmospheric neutrino evolution, see~\cite{Arguelles:2022hrt}. 

\begin{figure}[b]
    \centering
    \includegraphics[width=\textwidth]{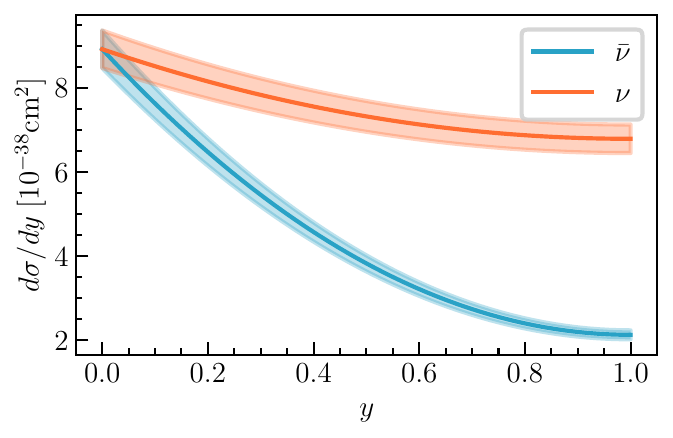}
    \caption{Differential neutrino charge current cross-section for the  DIS regime. The shaded region corresponds to the $1\sigma$ uncertainty region included for DIS in this analysis.}
    \label{fig:xsection}
\end{figure}

The different oscillation patterns between neutrinos and anti-neutrinos suggest that the separation of both particle types in the event basis is the best way to explore the aforementioned parameters.
In accelerator experiments, this is done by running the experiment in both the neutrino and anti-neutrino modes, while in the case of atmospheric neutrinos, the flux contains both neutrino types.
Therefore, we look for an alternative way to discriminate between neutrino- and anti-neutrino-type events. 

\begin{figure*}[t!]
   \centering
    \includegraphics[width=\textwidth]{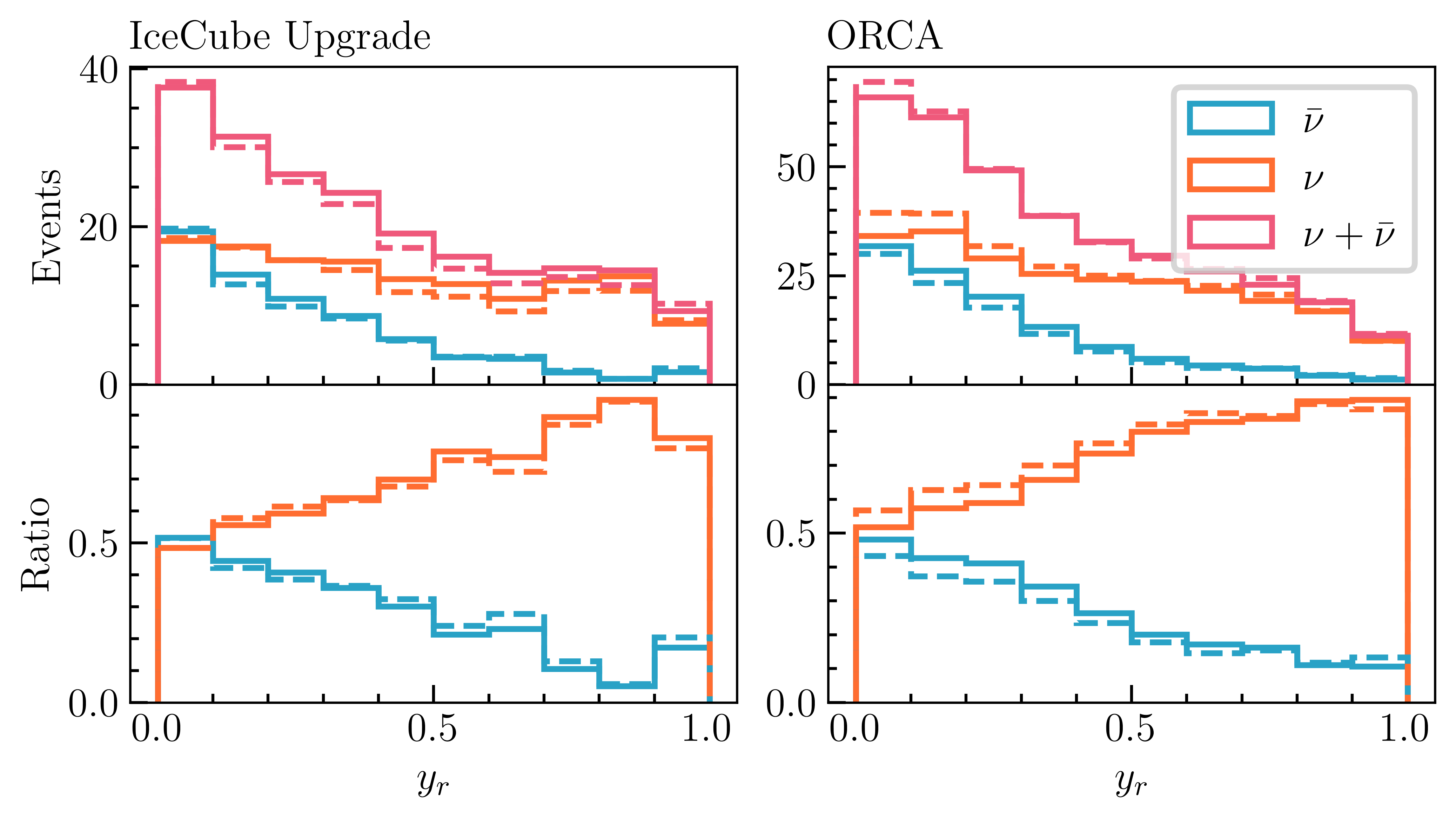}
    \caption{Histogram of events in IceCube Upgrade and ORCA as a function of $y_{r}$ for a bin in $E_{r}\in [5.0, 6.3]$~GeV and $\cos\theta_{r}\in[-0.8,-0.6]$. The solid line corresponds to Normal Ordering and the dashed line to Inverted Ordering. In the lower panel, we display the ratio between the neutrinos (orange) and anti-neutrinos (blue) and the total number of events for both normal and inverted ordering.}
    \label{fig:ORCA_events}
\end{figure*}

Following a neutrino's charged-current (CC) interaction with a nucleon ($N$), the neutrino energy is split between the leptonic ($l_{\alpha}$) and hadronic ($h$) currents, $\nu_{\alpha} + N \rightarrow l_{\alpha} + h$.
The $V-A$ structure of the weak tensor in the case of the neutrino interaction results into a different cross-section for neutrinos and anti-neutrinos~\cite{Ribordy:2013xea}, when neutrinos interact primarly with valence quarks, which is the case at the relevant energies.
Considering just the case where the neutrinos interact via Deep-Inelastic Scattering (DIS), the relevant component above 3~GeV, and following the notation in~\cite{Halzen:1984mc}, the neutrino cross-section can be written in terms of the inelasticity ($y = 1 - E_{l}/E_{\nu}$), where $E_{l}$ is the energy of the outgoing lepton, and the Bjorken scaling variable ($x$) as 
\begin{align}
\frac{d\sigma^{CC}_{\nu}}{dydx} = \frac{G^2_{F} xs}{2\pi}\left( Q(x) + \bar{Q}(x)\times(1-y)^2\right),\\
\frac{d\sigma^{CC}_{\bar{\nu}}}{dydx} = \frac{G^2_{F} xs}{2\pi} \left(\bar{Q}(x) + Q(x) \times(1-y)^2\right),
\end{align}
where $G_{F}$ is the Fermi constant and $s$ is the square of the center-of-mass energy.
The symbols $Q(x)$ and $\bar{Q}(x)$ corresponds to the sum of all the parton distribution functions (PDFs) for quarks and anti-quarks that contribute to the nucleons.
To explore the dependence of the neutrino cross-section on the inelasticity, we have integrated the double differential cross-section over $x$ within the kinematic allowed region, and using the PDF4LHC21 set~\cite{PDF4LHCWorkingGroup:2022cjn} PDFs set.
We find an almost uniform energy distribution of the outgoing lepton in the case of the neutrino interaction, as shown in \Cref{fig:xsection}.
In the case of the anti-neutrino interaction, most of the energy of the incoming neutrino is carried out by the outgoing lepton.
Therefore, it is possible to get a large neutrino-anti-neutrino separation for large values of $y$. 

Although we have restricted the discussion in this section to the DIS interaction, all the results that are presented in this analysis are based on simulations that includes all the interaction channels, as described in~\cite{Arguelles:2022hrt}.

%\PFM{Our simulations consider all interaction channels as described in \cite{Arguelles:2022hrt}, we take DIS as illustrating example.}

\textbf{\emph{Experiments and Methods---}} %\label{sec:experiments}
The IceCube Neutrino Observatory~\cite{Achterberg_2006} is an ice-Cherenkov neutrino detector located on average $2$~km below the surface at the geographic South Pole.
It consists of 5160 light sensors known as digital optical modules (DOMs) that allow it to detect neutrino interactions above $\sim 10$~GeV.
Depending on the type of particle propagating through the ice, an event will correspond to one of two possible morphologies, namely, \textit{tracks}, coming from the propagation of muons, and \textit{cascades}, coming from the propagation of electrons, taus, and/or hadronic or electromagnetic cascades.
In the near future, a detector upgrade~\cite{Ishihara:2019aao,Stuttard:2020zsj} that will consist in the deployment of additional strings allowing to lower the energy threshold to a few GeV.

We further consider the ORCA detector, which is part of the KM3NeT water-Cherenkov neutrino telescope currentlty under construction in the Mediterranean Sea~\cite{Adrian_Martinez_2016}.
As in the case of IceCube, ORCA also identifies tracks and cascades as possible event morphologies, but have also developed a third sample, namely, the \textit{intermediate}, for events that cannot be clearly identified as part of the former two.
For the purposes of our analysis, we use the open-access Monte Carlo simulation of ORCA developed in~\cite{Arguelles:2022hrt}, which is built as an extension of the open-access IceCube-Upgrade Monte Carlo release.

In both experiments, we compute the inelasticity for charged-current $\nu_{\mu}$ events which produce an outgoing muon, reconstructed as a track, and a hadronic shower, identified as a cascade.
 In terms of reconstructed quantities,
%The case of a $\nu_{\mu}$ charged current (CC) interaction with the ice produces an outgoing muon, seen as a track in the event reconstruction, and hadronic shower, identified as a cascade. 
%By reconstructing both of these morphology components we are able to infer the inelasticity of the event.
\begin{equation}
    y_r = \frac{E^{\text{casc}}_r}{E^{\text{casc}}_r + E^{\text{track}}_r}.
\end{equation}

Current oscillation analysis carried away by IceCube and ORCA use a two-dimensional histogram of the events in terms of the reconstructed energy and the direction for each morphological category.
To incorporate the inelasticity in the analysis, we modified the Monte Carlo simulations for IceCube-Upgrade~\cite{IceCube_Collaboration2020-md} and ORCA by adding a variable corresponding to the reconstructed inelasticity, $y_r$.
For every MC event reconstructed as a track, we generate a set of $N$ additional events, where the inelasticity is reconstructed based on the reconstructed energy for the track and the cascade.
For the main results of this work, we have assumed a Gaussian distribution with uncertainty of $\sigma_T = 20\%$ for tracks and $\sigma_C = 30\%$ for cascades.
For the main results of this work, we have assumed a Gaussian distribution with uncertainty of $\sigma_T = 20\%$ for tracks and $\sigma_C = 30\%$ for cascades~\cite{IceCubeCollaboration:2023wtb}.
For the purposes of this letter, we used the case of $N = 20$ for the IceCube simulation and $N = 10$ for ORCA~\footnote{We explored the sensitivity considering different values for $N$ between $N=10$ and $N=100$, finding no significant deviation of the results.}.
In addition to the binning scheme described in~\cite{Arguelles:2022hrt}, a third dimension is implemented for track events of both experiments, including $10$ bins for the reconstructed inelasticity.
An example of the event distribution we predict is shown in~\Cref{fig:ORCA_events}, where we have chosen one bin in zenith $\cos\theta_{r}\in [-0.8,-0.6]$ and energy $E_{r}\in[5.0, 6.3]$~GeV.
As anticipated from the previous discussion, for large values of $y$, the event distribution is primarily dominated by the neutrino sample.
In the case of anti-neutrinos, the event distribution is concentrated in the bins with small $y_{r}$.
The event distribution is depicted for both mass orderings, normal (solid), and inverted (dash).
For neutrinos, the mass ordering induce a event deviation which is almost uniform in $y_{r}$.
For the case of anti-neutrino, this deviation concentrates at lower $y_{r}$.

\textbf{\emph{Analysis and Results ---}}
%\label{sec:results}
We have investigated how the sensitivity to oscillation parameters improves with the inclusion of inelasticity in the analysis.
Through a combined analysis using currently publicly available IceCube-Upgrade and ORCA simulations, we have explored the sensitivity to the less constrained oscillation parameters—$\sin\theta_{23}$, mass ordering, and the CP-phase.
In this work, we have kept the solar parameters ($\Delta m^2_{21}$ and $\sin^2\theta_{12}$) and the reactor angle ($\sin^2\theta_{13}$) fixed at their best-fit values~\cite{Esteban:2020cvm}.
Regarding systematic uncertainties, we have taken into account uncertainties associated with the atmospheric neutrino flux, neutrino cross-section, and detector response.
These uncertainties have been included in the analysis in a manner similar to that presented in~\cite{Arguelles:2022hrt}.

\begin{figure}[t]
    \centering
    \includegraphics[width=\textwidth]{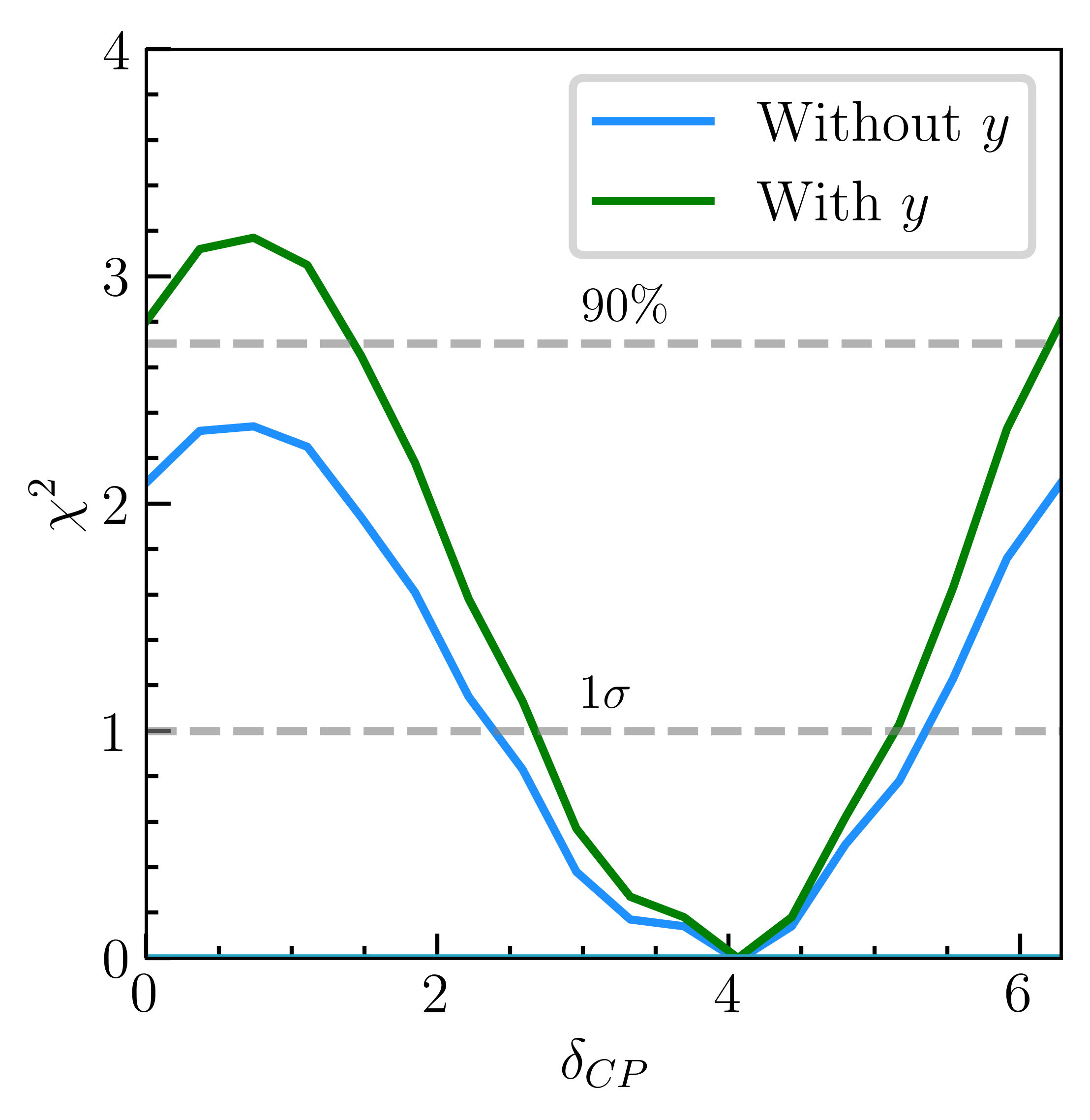}
    \caption{Comparison of the sensitivity from the combined analysis of Icecube-Upgrade (5 years) and ORCA (3 years) to $\delta_{CP}$ (solid lines) incorporating the $y_r$ binning (green) and the usual analysis (blue), assuming true normal ordering.}
    \label{fig:DCP}
\end{figure}

\begin{figure}[b]
    \centering
    \includegraphics[width=\textwidth]{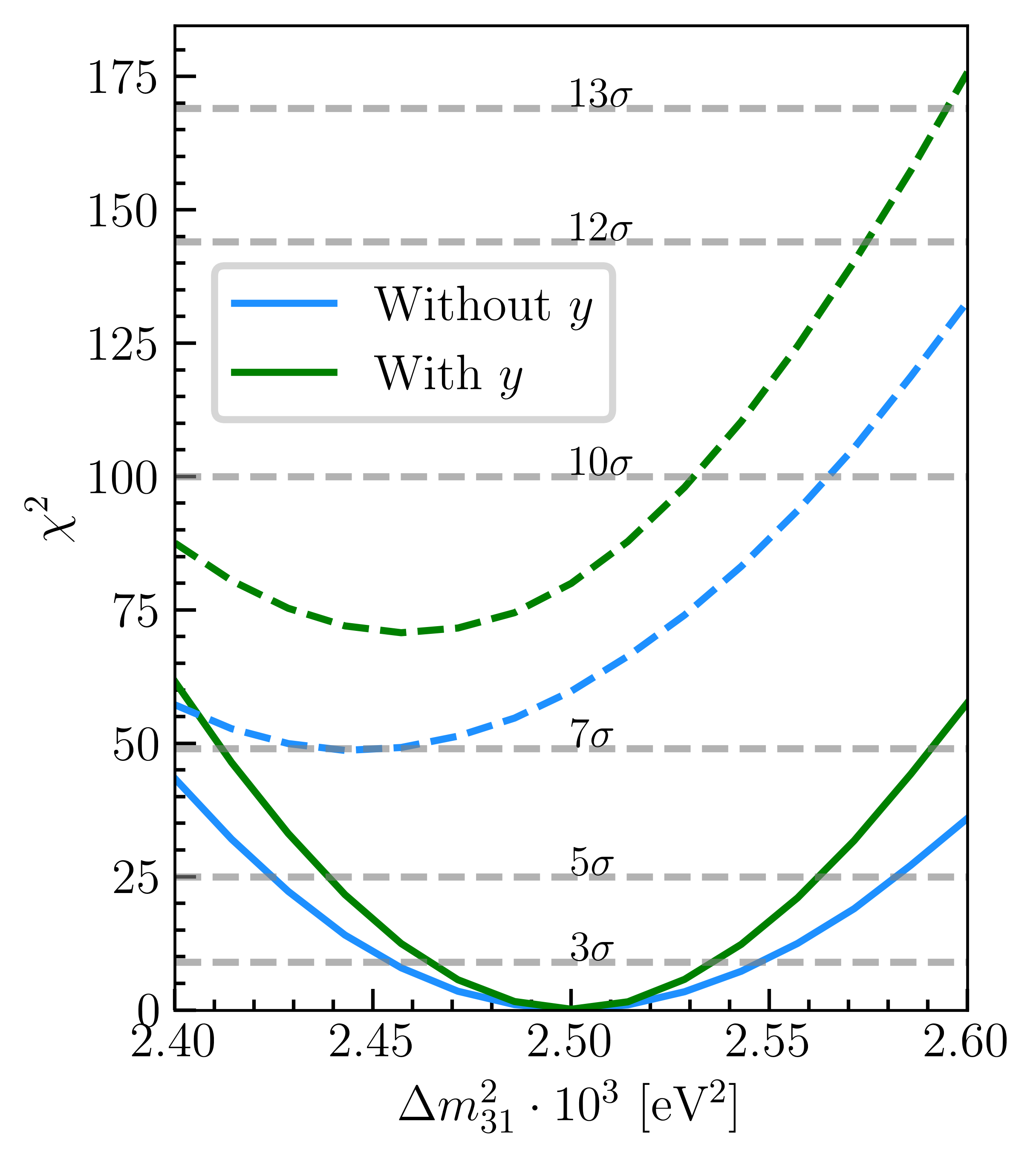}
    \caption{Comparison of the sensitivity from the combined analysis of Icecube-Upgrade (5 years) and ORCA (3 years) to $\Delta m^2_{31}$ (solid lines) incorporating the $y_r$ binning (green) and the usual analysis (blue), assuming true normal ordering. Dashed lines are the inverted ordering fit, showing the NMO sensitivity in units of $\chi^2$ as a function of $\Delta m^2_{31}$. In this analysis, we have profiled over $\sin\theta_{23}$ and $\delta_{CP}$, maintaining their best fit equal to the latest result from the global analyses.}
    \label{fig:Dm231}
\end{figure}

The main results of the combined analysis are illustrated in~\Cref{fig:DCP,fig:Dm231}.
The sensitivities depicted correspond to the combination of IceCube-Upgrade and ORCA, with exposures of 5 and 3 years, respectively.
In both figures, we assume normal ordering as the benchmark scenario.
Regarding $|\Delta m^2_{31}|$, we observed an improvement of more than $30\%$, achieving a precision below the percent level ($0.7\%$), as depicted in~\Cref{fig:Dm231} (solid lines).
However, for $\sin^2\theta_{23}$, which is influenced by the neutrino angular resolution, no improvement is observed.
For both parameters, profiling has been performed over $\delta_{CP}$ and the parameters not shown.

%The solid lines in each show the sensitivity to the specific parameter in units of $\chi^2$ as a function of possible parameter values. 

The sensitivity to the ordering is depicted in~\Cref{fig:Dm231} by the dashed lines.
We fit the event distribution assuming inverted ordering to the normal ordering scenario.
The combination of IceCube~Upgrade and ORCA will enable us to predict a $7\sigma$ exclusion of the inverted ordering without including the inelasticity, as shown in~\cite{Arguelles:2022hrt}.
With the inclusion of inelasticity in the analysis, the sensitivity increases to $\sim8.4\sigma$.
When considering each experiment separately, IceCube~Upgrade can reach $\sim 5\sigma$ in 3.5 years, while ORCA can do it in  2.5 years.

Finally, in the context of the CP-violating phase, although it does not have a significant impact on the muon disappearance channel, the inclusion of inelasticity in the analysis increases the resolution of $\delta_{CP}$ by $\sim 15\%$.
%and enables us to achieve more than $90\%$ exclusion for a large fraction of the parameter space.

%Additionally, the dashed lines of \Cref{fig:Dm231} show the fit of inverted ordering, thereby showing the NMO sensitivity. The green lines correspond to sensitivities obtained via an analysis incorporating the inelasticity, for which we used 10 bins in $y_r$, while the blue lines show the usual analysis. We can see how incorporating the inelasticity leads to significant improvements in the sensitivity to both the NMO and the value of $\Delta m^2_{31}$ and $\delta_{CP}$. 

Furthermore, to assess the resilience of our results, we investigated how the new sensitivity changes under limitations related to energy reconstruction and the possible misclassification of events with large inelasticity.
These tests confirmed the robustness of our method to these potential errors; refer to the Appendix for detailed information.

% \begin{figure*}
%     \centering
%     \includegraphics[width=0.9\textwidth]{figures/Improvement_merged.png}
%     \caption{Comparison of the sensitivity from the combined analysis of Icecube-Upgrade (5 years) and ORCA (3 years) to $\Delta m^2_{31}$ and $\delta_{CP}$ (solid lines) incorporating the $y_r$ binning (green) and the usual analysis (blue). We assume true normal ordering, solid lines are the inverted ordering fit and solid lines show the sensitivity in units of $\chi^2$ as a function of $\Delta m^2_{31}$.}
%     \label{fig:IC_ORCA_improvement}
% \end{figure*}

\textbf{\emph{Conclusion ---}}
%\label{sec:conclusion}
%In this letter, we have shown the success of a modification in the analysis of events in water(ice)-Cherenkov neutrino detectors in improving their sensitivity to the neutrino mass ordering and the values of $\delta_{CP}$ and $\Delta m^2_{31}$. This modification consists of incorporating an event's inelasticity in the analysis, which can be done by considering $\nu_\mu$ CC interactions and separately reconstructing the resulting track and cascade energies, thereby being able to infer the inelasticity. Specifically, we considered the case of the IceCube-Upgrade and ORCA detectors and performed a combined sensitivity analysis using MC simulations of their events.
In this letter, we introduced a novel approach to the oscillation analysis of the atmospheric neutrino data suitable for the upcoming IceCube-Upgrade and ORCA experiments.
We motivate and demonstrate that introducing the information of the reconstructed inelasticity of track events has the potential to discern neutrinos and anti-neutrinos in the few GeV region, thus impacting the sensitivity of the relevant oscillation parameters, namely the neutrino mass ordering, the squared mass difference and the CP-phase.
This results motivate the development of reconstruction techniques that can infer the inelasticity for sub-100~GeV energies.

This work builds up the results from~\cite{Arguelles:2022hrt} showing the relevant role of atmospheric neutrinos in unequivocal measuring the neutrino mass ordering before the end of the decade and constraining the allowed values for the remaining oscillation parameters independently from the long-baseline programs.

%Accelerating the quest for the precise measurement of the lepton CP-phase.

\textbf{\emph{Acknoledments ---}}
SG acknowledges support from the Harvard College Research Program in the fall of 2023.
CAA are supported by the Faculty of Arts and Sciences of Harvard University, the National Science Foundation, IAIFI, and the David \& Lucile Packard Foundation.
CAA and IMS were supported by the Alfred P. Sloan Foundation for part of this work.
MJ is supported by the National Science Foundation and the Faculty of Arts and Sciences of Harvard University. IMS is supported by STFC grant ST/T001011/1.

\nocite{*}
\newpage
\bibliography{main}% Produces the bibliography via BibTeX.
%%%%%% SUPPLEMENTAL MATERIAL STARTS HERE

\clearpage
\newpage

\onecolumngrid
\appendix

\ifx \standalonesupplemental\undefined
\setcounter{page}{1}

%\newcounter{SIfig}
%\setcounter{SIfig}{1}
\setcounter{figure}{0}

\setcounter{table}{0}
\setcounter{equation}{0}
\fi
\renewcommand{\thepage}{Supplemental Methods and Tables -- S\arabic{page}}

\renewcommand{\figurename}{SUPPL. FIG.}

\renewcommand{\tablename}{SUPPL. TABLE}
\renewcommand{\theequation}{A\arabic{equation}}

\newcounter{SIfig}
\renewcommand{\theSIfig}{SUPPL. FIG. \arabic{SIfig} }

\section{Uncertainty Associated to the Inelasticity}\label{appx:error}
In order to test the robustness of our results to different ways of modifying an experiment's Monte Carlo simulation when including the reconstructed inelasticity, $y_r$, we explored the NMO sensitivity obtained when $y_r$ is drawn directly from a Gaussian distribution, instead of determined via the method described above relying on two separate Gaussian distributions for the reconstructed cascade and track energies. We describe this method in the current appendix, followed by its resulting sensitivity in \Cref{appx:impact}. As above, for every $\nu_\mu$ CC interaction event having a true inelasticity of $y$, we create $N$ additional events with an additional variable, $y_r$. Instead of computing $y_r$ from reconstructed energies, we assume that \begin{equation}
    y_{r}\sim\mathcal N(y, \sigma_y),
\end{equation}
where the standard deviation, $\sigma_y$, is given by

\begin{equation}
    \sigma_y = y\sqrt{(1-y)^2\sigma_L^2 + y^2\sigma_H^2}.
\end{equation}

Here $\sigma_L$ and $\sigma_H$ represent the magnitude of the error in the reconstruction of $y$ for small and large values of $y$, respectively. To see why, notice that the factor multiplying $y$ in the above approaches $\sigma_L$ as $y\to 0$ and $\sigma_H$ as $y\to 1$. We experimented with different values of these parameters and decided to use $(\sigma_L, \sigma_H) = (0.2, 0.3)$, for this would reproduce the errors used in the main analysis in the limits of low and high inelasticity, respectively. During the modification of an experiment's MC simulation, in the case that a draw from this distribution is outside the physical range $[0, 1]$, we simply continue taking draws from this distribution until we obtain a $y_r$ value in this physical range. \\

\begin{figure}[h]
    \centering
    \includegraphics[scale=0.8]{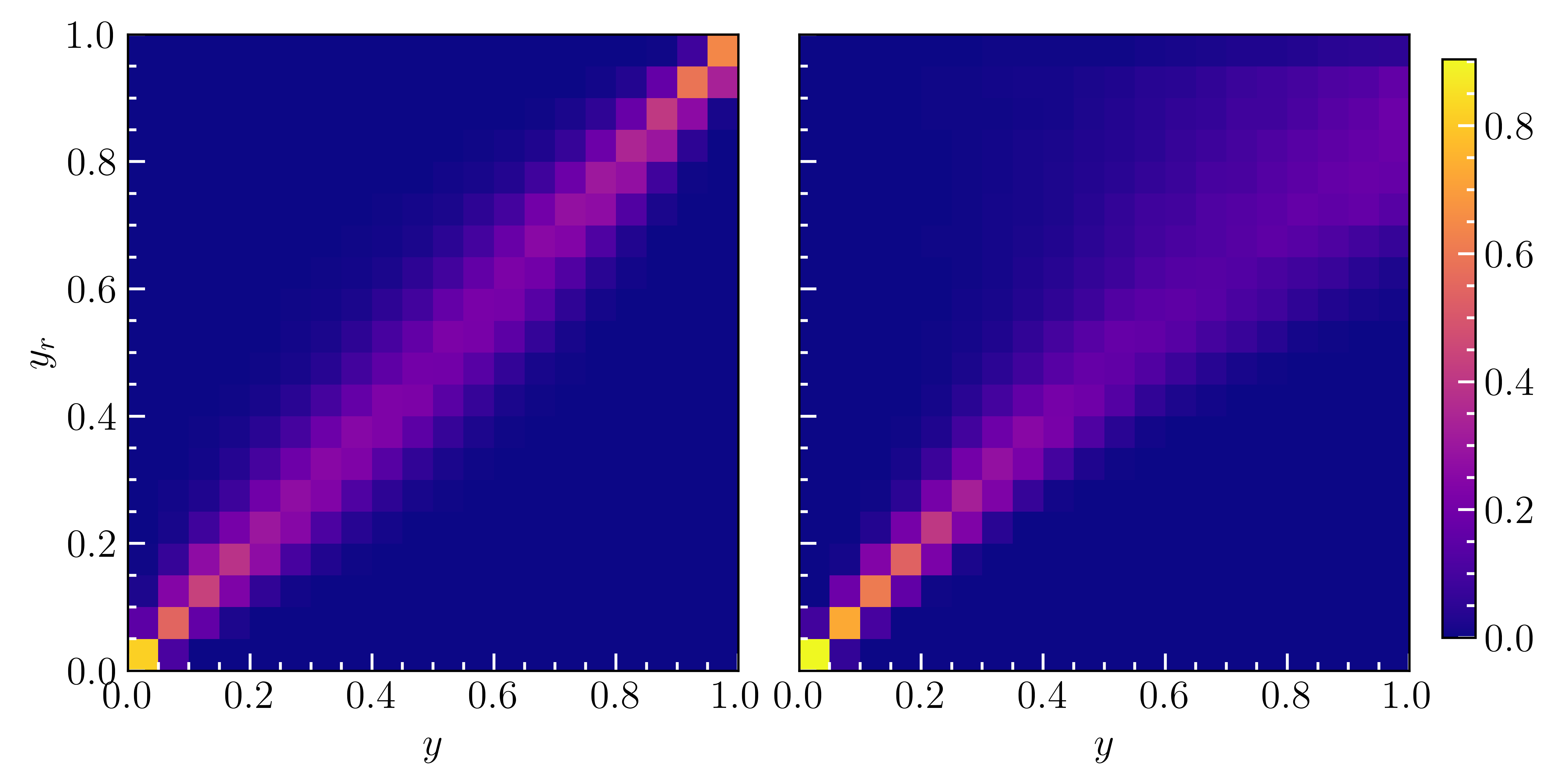}
    \caption{Distribution of the reconstructed inelasticity, $y_r$, and the corresponding true value, $y$, in IceCube-Upgrade. The left plot corresponds to the modification of the MC explained in the main paper, where we draw samples from a Gaussian distribution for each of the reconstructed track and cascade energy, and calculate $y_r$ from them. The right plot corresponds to the case where $y_r$ is drawn directly from a Gaussian distribution, $y_r\sim\mathcal N(y, \sigma_y)$.}
    \refstepcounter{SIfig}\label{fig:appA}
\end{figure}

The distribution of $y_r$ obtained using this method is plotted in \ref{fig:appA}.

\section{Impact of Inelasticity in the Oscillation Analysis}\label{appx:impact}
\begin{figure}[h]
    \centering
    \includegraphics[scale=0.8]{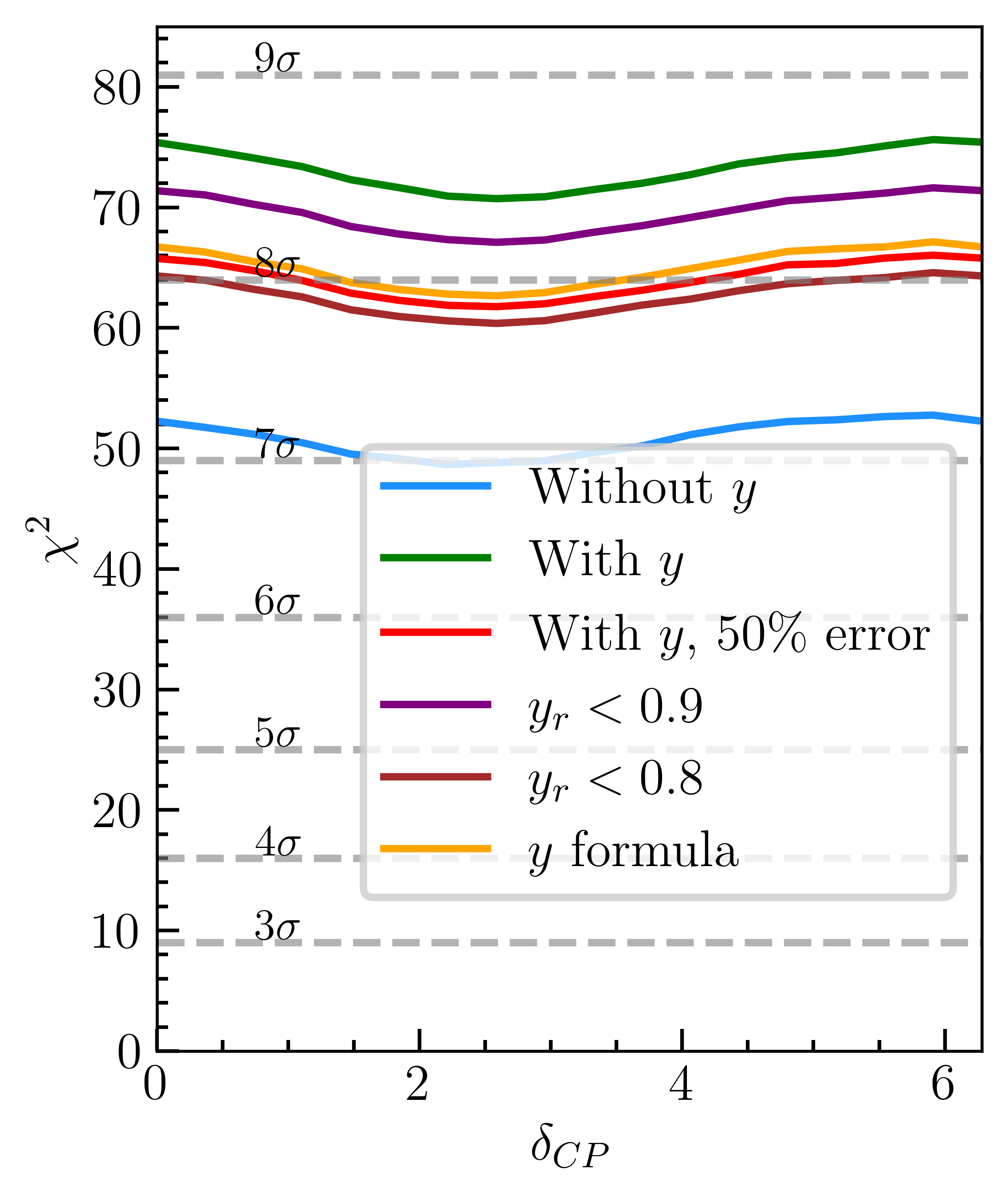}
    \caption{Combined NMO sensitivity of IceCube-Upgrade (5 yrs) and ORCA (3 yrs). The green line shows the sensitivity obtained by including the $y$ and the blue is the usual analysis. The red line  shows the case of poor energy reconstruction when inferring $y_r$, assuming a 50\% error for both the track and cascade reconstructed energies. The purple and maroon lines show what happens when we only include the bins where $y_r < 0.9$ and $y_r < 0.8$, respectively, corresponding to the case where the high-inelasticity events are misclassified. Lastly, the orange line shows the sensitivity obtained when $y_r$ is drawn directly from a Gaussian centered at each true $y$, as explained in \Cref{appx:error}.}
    \refstepcounter{SIfig}\label{fig:appB}
\end{figure}

When assessing whether or not incorporating the inelasticity in the event analysis is worthwhile, two natural questions to ask are: (1) how robust is this method with respect to errors in the energy reconstruction of the track and cascade, and (2) is the resulting improvement in sensitivity coming only from bins with high $y$ value, which could be misidentified as cascades by the reconstruction? Here we show how this proposed method is indeed robust to poor energy reconstruction and to misidentifications of events with large $y$ as purely cascades.
In \ref{fig:appB}, we show the impact that a poor reconstruction of the separate energies of the track and cascade would have on the resulting sensitivity to the mass ordering. We can see how, even making the very conservative assumption of a 50\% error in the reconstruction of both the track's and the cascade's energy, there is still substantial improvement in the NMO sensitivity compared to the analysis that does not incorporate $y$ at all. Therefore, our method is robust to errors in energy reconstruction.

Furthermore, \ref{fig:appB} shows what happens to the NMO sensitivity when we consider events with $y_r < 0.9$ and $y_r < 0.8$, that is, when we exclude high-$y_r$ bins.
This captures what would happen if all such events were incorrectly classified during the reconstruction.
As in the case of poor energy reconstruction, the sensitivity decreases when we ignore this bin, but the overall value is still higher compared to the analysis that does not incorporate $y$. Thus, the improvement is robust to cases where the event morphology is classified incorrectly.

\end{document}